\documentclass[pra,twocolumn,showpacs,preprintnumbers,amsmath,amssymb]{revtex4}
\usepackage{epsfig}
\usepackage{color}
\usepackage{graphicx}
\usepackage{dcolumn}
\usepackage{bm}
\newcommand{\bra}[1]{\langle#1|}
\newcommand{\ket}[1]{|#1\rangle}

\begin{document}
\title{Entanglement degradation in the solid state: interplay of adiabatic and
quantum noise}
\author{B. Bellomo,$^1$ G. Compagno,$^1$ A. D'Arrigo,$^2$ G. Falci,$^2$ R. Lo Franco,$^1$ and E. Paladino$^2$}
\affiliation{$^1$CNISM \& Dipartimento di Scienze Fisiche ed Astronomiche, Universit\`a di Palermo,
via Archirafi 36, 90123 Palermo, Italy\\
$^2$ Dipartimento di Metodologie Fisiche e Chimiche,
Universit\`a di Catania, viale A. Doria 6, 95125 Catania, Italy
\& CNR-IMM MATIS}

\date{\today}

\begin{abstract}
We study entanglement degradation of two non-interacting qubits subject to independent
baths with broadband spectra typical of solid state nanodevices. We obtain the analytic
form of the concurrence in the presence of adiabatic noise for classes of entangled initial
states presently achievable in experiments. We find that adiabatic (low frequency) noise
affects entanglement reduction analogously to pure dephasing noise. Due to quantum (high
frequency) noise, entanglement is totally lost in a state-dependent finite time. The
possibility to implement on-chip both local and entangling operations is briefly discussed.
\end{abstract}


\maketitle
Over the last decade, considerable progress has been made towards
the implementation of an electrically controlled solid-state
quantum computer. In particular, superconducting
high-fidelity~\cite{single-super,highfid} single qubit gates with coherence times
of about $1 \mu s$ are nowadays available~\cite{schreier,vion}.
The possibility to implement two-qubit logic gates has been
proved in different laboratories~\cite{coupled-exp}
and Bell states preparation has been demonstrated~\cite{bellgeneration}.
Recently, highly entangled states with concurrence up to $94$
per cent have been generated ``on demand'' in a circuit quantum
electrodynamic architecture, opening the way to the implementation of quantum algorithms with a
superconducting quantum processor~\cite{dicarlo}.

In order to achieve the high performances required to overcome classical
processors, it is important to establish how long a sufficient
degree of entanglement can be maintained in noisy nanocircuits.
Implications are the possibility to store entangled states
in solid-state memories and entanglement preservation
during local operations in quantum algorithms~\cite{Nielsen,yu2009Science}.
Solid state noise may represent a serious limitation towards this
goal. Superconducting nanodevices are usually affected by  broadband noise.
Typical power spectra display a $1/f$ low-frequency  behavior
followed by a white or ohmic flank~\cite{ithier,nak-spectrum}.

The effects on single-qubit gates of low- and high-frequency noise components
are quite different. Adiabatic (low frequency) noise typically leads to power-law
decay, quantum (high frequency) noise to exponential behavior~\cite{falci2005PRL,ithier}.
On the other hand, disentanglement may markedly differ from the single qubit decoherence.
For instance, at zero temperature, single qubit exponential decay due to a Markovian bath
contrasts to finite-time bipartite entanglement degradation, known as ``Entanglement
Sudden Death'' (ESD)~\cite{yu2006PRL}. In structured environments non-Markovian noise 
appears to be more fundamental~\cite{bellomo2007PRL,bellomo2008PRA}. Extending the 
current research on ESD into physically relevant non-Markovian situations remains a 
challenge~\cite{yu2009Science}. The analysis of entanglement degradation under the simultaneous 
presence of adiabatic and quantum noise places in this context and it may provide new 
insights to the exploitation of solid-state nano-devices for quantum information.

In this Communication we address these issues. We consider two non-interacting qubits subject
to independent baths with typical solid-state broadband spectra~\cite{NJP-special}.
Entanglement is quantified by the concurrence~\cite{wootters1998PRL}, which is evaluated in analytic
form in the presence of adiabatic noise for classes of entangled initial states
presently achievable in experiments.
We find that adiabatic noise has the same qualitative effect
of pure dephasing noise and no ESD occurs for pure initial states.
However, due to the interplay with quantum noise, entanglement is lost in a finite time
which depends on the initial entangled state.
We comment on the sensitivity to experimental imperfections and qubits operating point.
The possibility to implement on-chip both local and entangling
operations is briefly discussed.

{\em Model and evolved two-qubit density matrix --}
The system, formed by uncoupled qubits $A$ and $B$  affected by
independent noise sources, is modeled by
$H_\textrm{tot}=H_A+H_B$. Each qubit
is an-isotropically coupled to a noise source
\begin{equation}
\label{totalHamiltonian}
H_\alpha = H_{Q, \alpha} -\frac{1}{2} \hat{X}_\alpha \sigma_{z,\alpha},
\quad
H_{Q, \alpha} = -\frac{1}{2}\vec{\Omega}_\alpha\cdot\vec{\sigma}_\alpha \,.
\end{equation}
Here $\hat{X}_\alpha$, $\alpha=A,B$, are collective environmental variables whose power spectra
are $1/f$,  $f \in [\gamma_m, \gamma_M]$ and white or ohmic
at $f \geq \tilde f$, where $\tilde f \leq \Omega_\alpha$ ($\hbar =1$)~\cite{ithier,nak-spectrum}.
According to a standard model, noise with $1/f$ spectrum can be originated from an ensemble of
bistable fluctuators with switching rates $\gamma$ distributed as $1/\gamma$~\cite{weissman1988RMP}.
The physical origin of these fluctuations depends on the specific setup.
Both the operating point (the angle $\theta_\alpha$ between $z$ and $\vec{\Omega}_\alpha$) and the
splittings $\Omega_\alpha$ are tunable.
By operating each qubit at the ``optimal point'', $\theta_\alpha=\pi/2$, partial reduction of defocusing
may be achieved~\cite{vion,ithier}.

The two-qubit Density Matrix (DM) elements are evaluated in the computational basis
 $\mathcal{B}=\{\ket{0}\equiv\ket{00},\ket{1}\equiv\ket{01}, \ket{2}\equiv\ket{10},
\ket{3}\equiv\ket{11}\}$, where $H_{Q,\alpha}\ket{0}_\alpha=-\frac{\Omega_\alpha}{2}\ket{0}_\alpha$,
$H_{Q,\alpha}\ket{1}_\alpha=\frac{\Omega_\alpha}{2}\ket{1}_\alpha$. Since each ``qubit+environment''
evolves independently, the time evolution operator of the composite system is the tensor product
of terms corresponding to the two parts. Thus, once the
single-qubit dynamics, expressed by the DM elements
$\rho^{A}_{ii'}(t)=\sum_{ll'}A_{ii'}^{ll'}(t)\rho^{A}_{ll'}(0)$,
$\rho^{B}_{jj'}(t)=\sum_{mm'}B_{jj'}^{mm'}(t)\rho^{B}_{mm'}(0)$,
is solved,
the evolved two-qubit DM can be evaluated as~\cite{bellomo2007PRL}
\begin{equation}\label{totalevo}
\langle i j | \rho(t) |i' j' \rangle =
\sum_{ll',mm'}A_{ii'}^{ll'}(t)B_{jj'}^{mm'}(t)
\langle l j | \rho(0) | l' j' \rangle,
\end{equation}
where indexes take values $0,1$.
We consider extended Werner-like (EWL) two-qubit initial states
\begin{equation}\label{EWLstates}
    \hat{\rho}^\Phi=r \ket{\Phi}\bra{\Phi}+\frac{1-r}{4}I_4,\quad
    \hat{\rho}^\Psi=r \ket{\Psi}\bra{\Psi}+\frac{1-r}{4}I_4,
\end{equation}
whose pure parts are the one/two-excitations Bell-like states
$\ket{\Phi}= a \ket{01}+ b \ket{10}, \quad
\ket{\Psi}= a \ket{00} + b \ket{11}$, where $|a|^2+|b|^2=1$.
The purity parameter $r$ quantifies the mixedness and $a$
sets the degree of entanglement of the initial state.
In the experiment of Ref.~\cite{dicarlo} entangled states with purity $\approx 0.87$ and
fidelity to ideal Bell states $\approx 0.90$ have been generated.
These states may be approximately described as EWL states with
$r_{exp} \approx 0.91$.

For EWL states, the density matrix in the computational basis is non-vanishing only along the
diagonal and anti-diagonal (X form)~\cite{yu2006PRL}; in this system this structure is maintained
at $t \geq 0$.
The entangled two-qubit dynamics is quantified by concurrence, $C(t)$ ($C=0$ for separable
states, $C=1$ for maximally entangled states)~\cite{wootters1998PRL}. For X states $C_\rho^X(t)=2\mathrm{max}\{0,K_1(t),K_2(t)\}$ \cite{yu2007QIC},
\begin{eqnarray}\label{concxstate}
\label{K1}
K_1(t)&=&|\rho_{12}(t)|-\sqrt{\rho_{00}(t)\rho_{33}(t)},
\\
\label{K2}
K_2(t)&=&|\rho_{03}(t)|-\sqrt{\rho_{11}(t)\rho_{22}(t)}.
\end{eqnarray}
The initial value of the concurrence is equal for both the EWL states of
Eq.~(\ref{EWLstates}) and reads
$C_\rho^{\Phi}(0)=C_\rho^{\Psi}(0)=2\mathrm{max}\{0,(|a \, b| +1/4)r-1/4\}$. Initial states are
thus entangled provided that $r>r^\ast=(1+4 |a \, b|)^{-1}$.

Elements of the two-qubit DM in the basis $\mathcal{B}$ will be obtained via Eq.~(\ref{totalevo}).
To solve the single-qubit dissipative dynamics we apply the multi-stage elimination approach
 introduced in Ref.~\cite{falci2005PRL}.
Effects of low- and high-frequency components
of the noise are separated  by putting,
 $\hat{X}_\alpha \to X_\alpha(t) + \hat X_\alpha^f$.
Stochastic variables $X_\alpha(t)$ describe low-frequency
($1/f$) noise, and can be treated in the adiabatic and
longitudinal approximation. High-frequency
($\omega \sim \Omega_\alpha$) fluctuations $\hat X_\alpha^f$
are modeled by a Markovian bath and mainly determine
spontaneous decay. Therefore, populations relax  due to quantum noise
($T_1$-type times), which also leads to secular dephasing
($T_2^*= 2 \, T_1$-type). Low-frequency noise provides
a defocusing mechanism determining further coherences decay.

\begin{figure}
{\includegraphics[width=4.2 cm, height=4 cm]{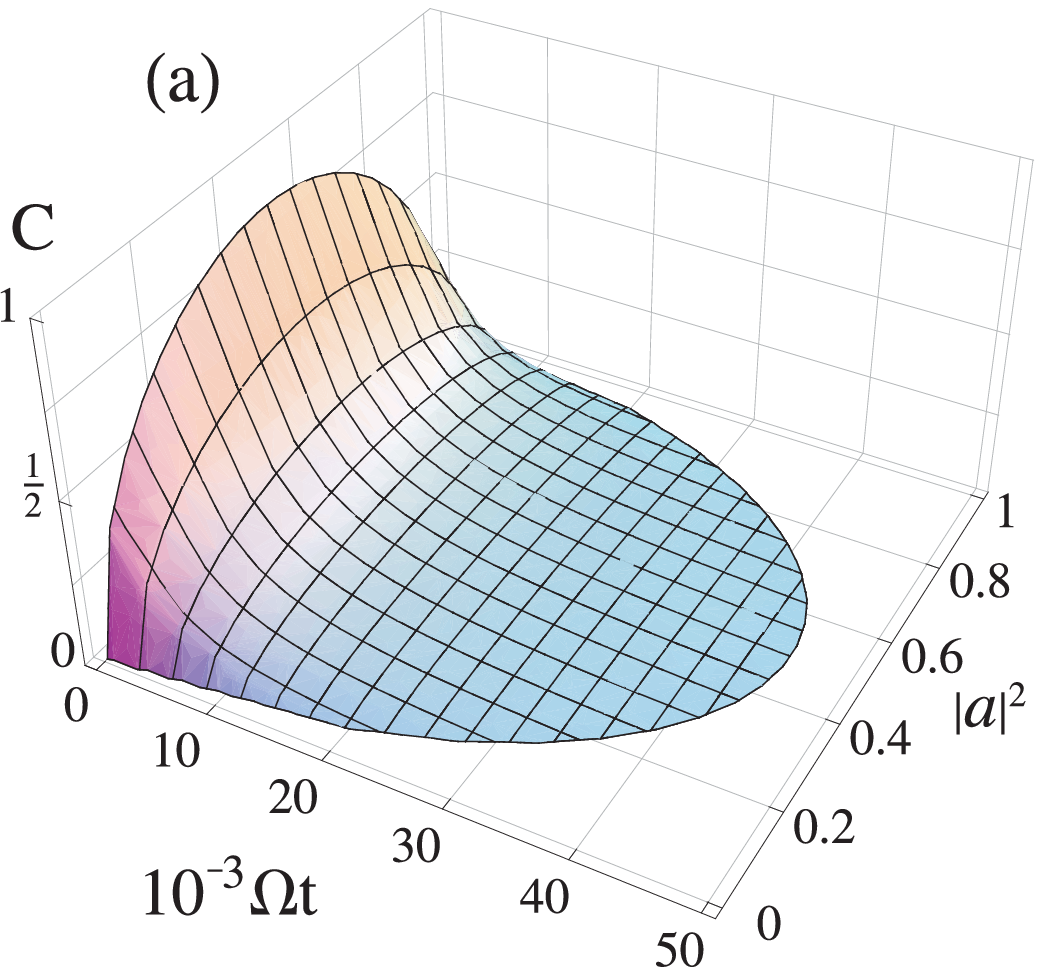}
\includegraphics[width=4.2 cm, height=4 cm]{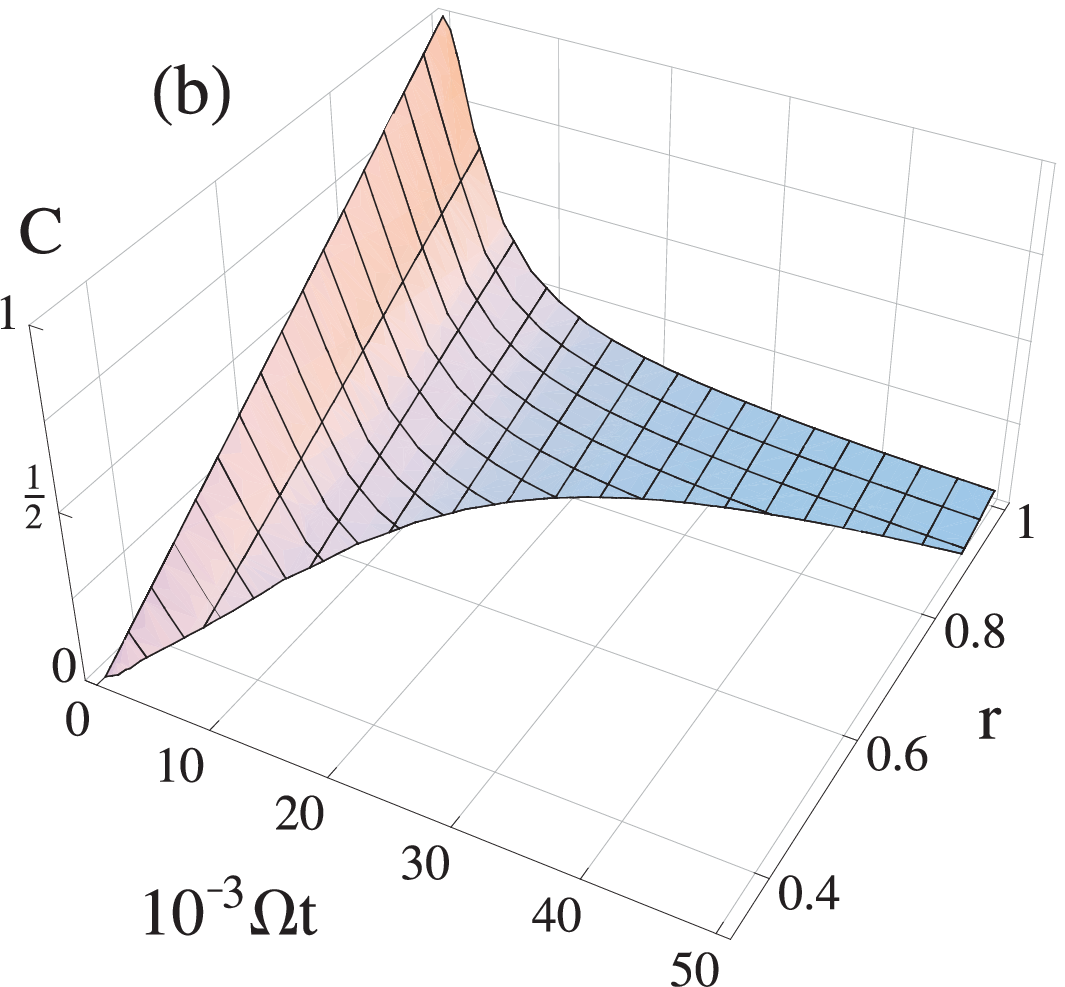}}
\caption{\label{fig1}\footnotesize (Color online)
 Concurrence (\ref{conc-adiabatic}) at $\theta=\pi/2$ and
$\Sigma/\Omega=0.02$. Panel (a) $C(t)$ as a function of
$|a|^2$ ($r=0.9$);
Panel (b) $C(t)$ vs $r$  ($a=1/\sqrt2$).}
\end{figure}
{\em Concurrence under adiabatic noise -- }
The effect of low-frequency noise components is obtained by
replacing $\hat X_\alpha \approx X_\alpha (t)$.
In the adiabatic and longitudinal approximation
single-qubit populations do not evolve in time,
$\rho^\alpha_{ii}(t)=\rho^\alpha_{ii}(0)$, where $i=0,1$.
The leading effect of low-frequency fluctuations
is defocusing, given within the static-path approximation (SPA),
$\rho_{ij}^\alpha(t) \approx \rho_{ij}^\alpha(0) z_\alpha(t)$ with
$z_\alpha(t) =
 \int d X_\alpha
P(X_\alpha) \exp[- i \omega_{ij}(X_\alpha) t]$.
It amounts to neglect effects of the slow fluctuators dynamics during time evolution.
In relevant situations the probability density can be assumed of Gaussian
form $P(X_\alpha)=\exp(- X_\alpha^2/2 \Sigma_{\alpha}^2)/\sqrt{2 \pi} \Sigma_{\alpha}$
and the coherences take the form reported in Ref.~\cite{falci2005PRL}.
The variance $\Sigma_\alpha$ can be estimated by independent measurement
of the amplitude of the $1/f$ power spectrum on the uncoupled qubits,
$S_\alpha^{1/f}(\omega) =\pi\Sigma_\alpha^2
[\ln(\gamma_M/\gamma_m) \, \omega]^{-1}$.

For the initial EWL states of Eq.~(\ref{EWLstates}), the concurrences reduce
respectively to $C_\rho^\Phi(t)=2\mathrm{max}\{0,K_1^\Phi(t)\}$ and
$C_\rho^\Psi(t)=2\mathrm{max}\{0,K_2^\Psi(t)\}$, where
\begin{eqnarray}\label{concurrences}
K_1^\Phi(t)&=& |\rho_{12}^\Phi(0)| |z_A(t)||z_B(t)|-\sqrt{\rho_{00}^\Phi(0)\rho_{33}^\Phi(0)},
\\
K_2^\Psi(t)&=& |\rho_{03}^\Psi(0)| |z_A(t)||z_B(t)|-\sqrt{\rho_{11}^\Psi(0)\rho_{22}^\Psi(0)} \, ,
\end{eqnarray}
give the same concurrence $C_\rho^\Phi(t)=C_\rho^\Psi(t) \equiv C(t)$
\begin{equation}
\label{conc-adiabatic}
C(t) = 2r  \,| a \, b |\,
 \Pi_\alpha \frac{
\exp{\{-\frac{1}{2}\frac{(c_\alpha \Sigma_\alpha t)^2}
{1+ (s_\alpha \Sigma_\alpha)^4 (t/\Omega_\alpha)^2}}\}}
{
{[1 + (s_\alpha \Sigma_\alpha)^4 (t /\Omega_\alpha )^2]}^{1/4}
}-\frac{1-r}{2}
\end{equation}
for times smaller than the adiabatic ESD time, $t_{ESD}^{ad}$ identified by
the condition $C(t_{ESD}^{ad})=0$.
In Eq.~(\ref{conc-adiabatic}) $c_\alpha=\cos\theta_\alpha$, $s_\alpha=\sin\theta_\alpha$.
Adiabatic noise in the longitudinal approximation leads to the same qualitative
behavior obtained for pure dephasing noise~\cite{yu2006PRL}, i.e. the
concurrence does not vanish at any finite time for a pure state, $r=1$~\cite{yu2007QIC}.
On the contrary, any small degree of mixedness leads to disentanglement at a finite time.
Note that $t_{ESD}^{ad}$ depends on the operating points $\theta_\alpha$. Here we report
the ESD times for identical qubits, $\Omega_\alpha=\Omega$, $\theta_\alpha=\theta$,
operating at the optimal point ($\theta=\pi/2$) and at pure dephasing ($\theta=0$)
\begin{eqnarray}
t_{ESD}^{ad} &=& \frac{\Omega}{\Sigma^2} \sqrt{ 16 |ab|^2 \frac{r^2}{(1-r)^2} -1},
\quad \left(\theta=\frac{\pi}{2}\right)
\label{tESD-opt} \\
t_{ESD}^{ad}&=&\frac{1}{\Sigma}\sqrt{\ln{\left [4|ab|\frac{r}{1-r}\right]}},\hspace{1.25 cm}
(\theta=0)
\end{eqnarray}
where we assumed both qubits affected by the same amplitude $1/f$ noise,
$\Sigma_\alpha \approx \Sigma$.
The dimensionless ESD time $\Omega t_{ESD}^{ad}$ is longer  at the optimal
point than for pure dephasing.
This is originated from the different (algebraic or exponential)
decay of the concurrence Eq.~(\ref{conc-adiabatic}) at the two operating points. 
This behavior, due to the non-Markovian nature of $1/f$ noise,
results in a different scaling of   $\Omega t_{ESD}^{ad}$ with $\Omega/\Sigma \gg 1$ and in
an algebraic or logarithmic (at $\theta=0$),dependence on $r$ and $a$. 
The degree of purity of the initial state, $r$, has a crucial quantitative role
on the ensuing entanglement maintenance. This is illustrated in Fig.~\ref{fig1} for
$1/f$ noise amplitude expected in single qubit experiments
$\Sigma/\Omega=0.02$~\cite{vion,ithier,falci2005PRL}. The dependence on the initial
degree of entanglement is instead smoother and symmetric around $|a|^2=1/2$.

In the multistage approach, quantum noise
($\omega \sim \Omega$) adds up to the defocusing channel leading
to extra exponential decay of the coherences and evolutions of the populations.
This last fact makes the concurrence of the two EWL states nonequivalent.
At finite temperature, quantum noise  leads to ESD even for initial Bell-like states
($r=1$)~\cite{bellomo2008PRA}. This is the main qualitative difference with
adiabatic noise. Here we discuss the interplay of adiabatic and quantum noise simultaneously
affecting nanodevices. We assume both qubits operate at the optimal point, where
the effect of adiabatic noise is reduced.
\begin{figure}[t!]
{\includegraphics[width= 6 cm, height=3.7 cm]{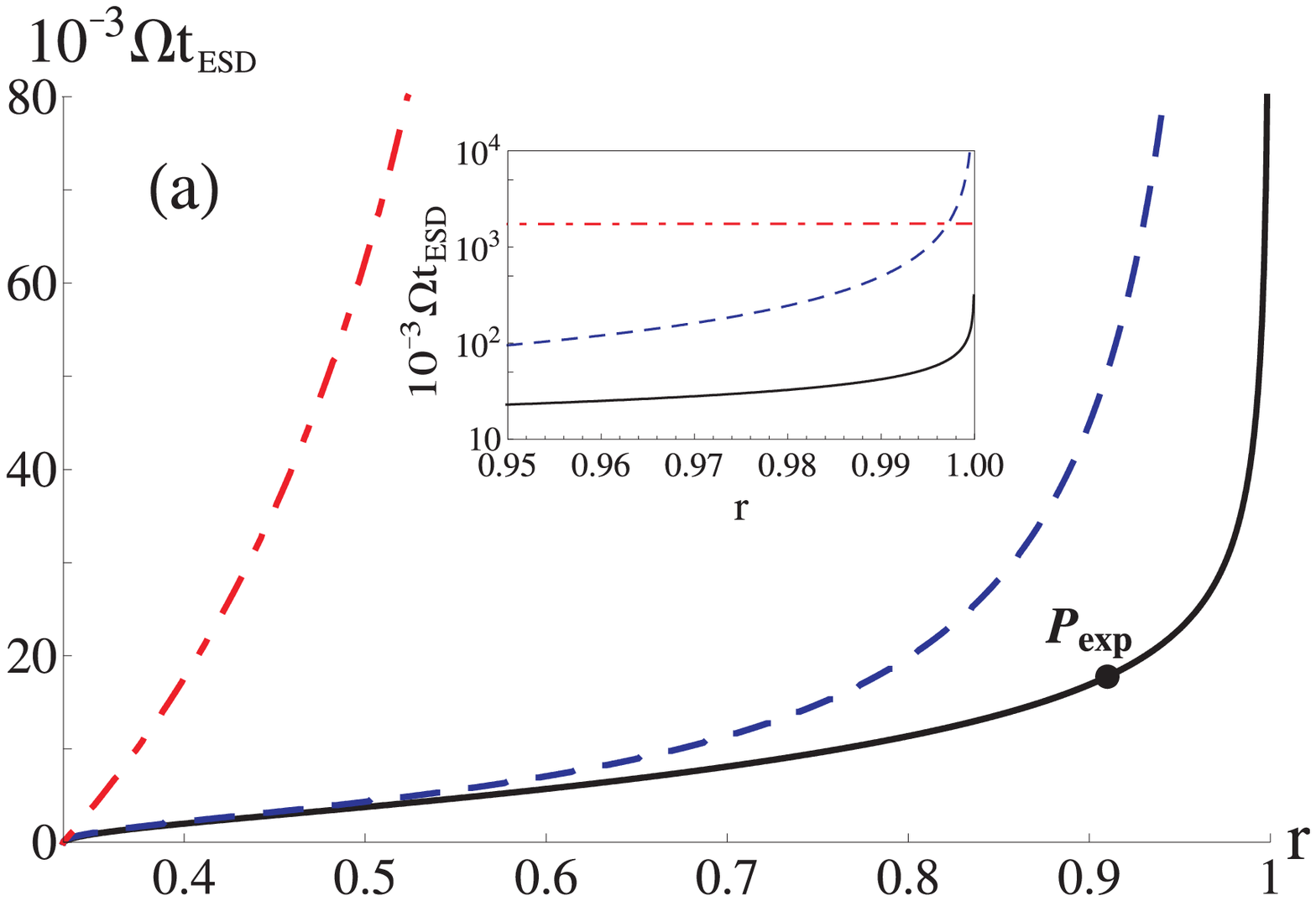}\vspace{0.5 cm}
\includegraphics[width=  6  cm, height=3.7 cm]{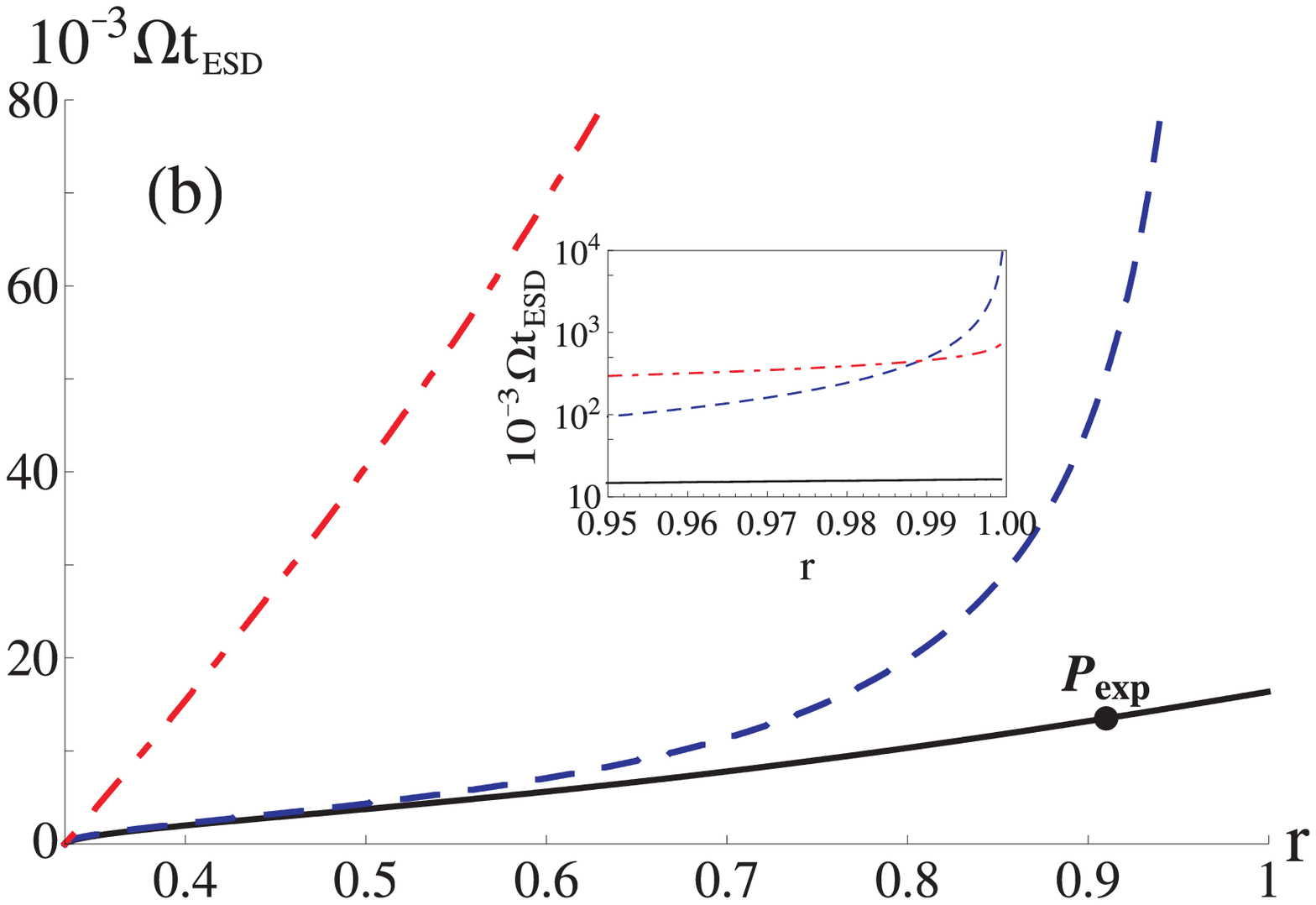}}
\caption{\footnotesize (Color online) Dependence of the ESD time on the purity $r$ ($a=1/\sqrt{2}$)
for initial state  $\hat{\rho}^\Phi$ (panel a), and $\hat{\rho}^\Psi$ (panel b). The
blue dashed curve is  $\Omega t_{ESD}^{ad}$, Eq.(\ref{tESD-opt}),
red dot-dashed curve is for quantum noise, black curve is the result of adiabatic and
quantum noise. Noise characteristics are $\Sigma=0.02\Omega$, $S_f(\omega)=2\times10^{6}$ s$^{-1}$.
In addition,  $\Omega=10^{11}$ rad/s,  $\theta=\pi/2$, $T=0.04$ K.
The inset zooms the region where $r\approx1$. The point $P_\mathrm{exp}$ corresponds
to $r_\mathrm{exp}\approx0.91$ where $\Omega t_\mathrm{ESD}\approx18\times10^3$ for
$\hat{\rho}^\Phi$ and $\Omega t_\mathrm{ESD}\approx14\times10^3$ for $\hat{\rho}^\Psi$.} \label{fig2}
\end{figure}

{\em Interplay of adiabatic and quantum noise  -- }
The concurrence is evaluated from Eqs.~(\ref{K1})~-~(\ref{K2})
with qubit $\alpha$ populations obtained from the Born-Markov master equation.
In the presence of white noise at frequencies $\omega \sim \Omega$,
they read
$
\rho_{ii}^\alpha(t) = (\rho_{ii}^\alpha(0)-\rho_{ii}^{\alpha \infty})\mathrm{e}^{-t/T_1} +\rho_{ii}^{\alpha \infty}
$
with  relaxation rate $T_1^{-1}=S_f(\Omega) /2$
and asymptotic population difference
$\rho_{11}^{\alpha \infty}-\rho_{00}^{\alpha \infty}=-\tanh(\Omega_\alpha/2k_BT)$.
The coherences acquire an additional exponential decaying factor,
$T_2^{-1} = T_1^{-1}/2$, and read
\begin{equation}
\label{eq:slow-fast}
 \rho_{01}^\alpha(t) \approx \rho_{01}^\alpha(0) \,
\mathrm{e}^{- i \Omega_\alpha t
  - \frac{1}{2} \,
\ln \big( 1 +
 \, \big(i \Omega_\alpha +  \frac{1}{T_1}  \big)
\,  \frac{\Sigma^2_\alpha  t}{ \Omega_\alpha^2}
\big) -\frac{t}{2T_1}} \, .
\end{equation}
The concurrence can be evaluated in analytic form. Expressions are quite lengthy
and here we exemplify the case of initial Bell states
($r=1$, $a=1/\sqrt 2$) and resonant qubits:
$C_\rho^\Phi(t)= 2 {\rm max}\{ 0,K_1^\Phi(t)\}, C_\rho^\Psi(t)=2{\rm max}\{ 0,K_2^\Psi(t)\}$
where
\begin{eqnarray}
&& K_1^\Phi(t)= \frac{1}{2}\frac{e^{-\frac{t}{T_1}}}{\sqrt{1+\Sigma^4(t/\Omega)^2}}-
\sqrt{\rho_{11}^\infty\rho_{00}^\infty} (1-e^{-\frac{t}{T_1}}) \nonumber \\
&&\times \sqrt{\left((\rho_{00}^\infty)^2+(\rho_{11}^\infty)^2\right)e^{-\frac{t}{T_1}}
+\rho_{00}^\infty\rho_{11}^\infty\left(1+e^{-\frac{2t}{T_1}}\right)}
\label{example1}
\\
&& K_2^\Psi(t) = \frac{1}{2} \frac{e^{-\frac{t}{T_1}}}{\sqrt{1+\Sigma^4(t/\Omega)^2}}
-  \frac{1}{2}(1-e^{-\frac{t}{T_1}})
\nonumber\\
&& \qquad \times
\left[ \left((\rho_{00}^\infty)^2+(\rho_{11}^\infty)^2\right)  e^{-\frac{t}{T_1}} +
 2 \rho_{00}^\infty \rho_{11}^\infty \right] \, .
\label{example2}
\end{eqnarray}
In general, due to the presence of quantum noise, entanglement is lost in a finite time for any
$r$ and $a$. A comparison of the ESD times in the presence of adiabatic noise, quantum noise and
their interplay is illustrated in Fig.(2) for white noise level expected from
single qubit experiments, $S_f(\omega)=2\times 10^6$s$^{-1}$ and for the two Werner states.
The overall ESD time for states $\hat\rho^\Phi(0)$ is longer than the one for $\hat\rho^\Psi(0)$.
This effect originates from relaxation processes due to quantum noise.
The two-excitation state $\ket{\Psi}$ can decay to a
one-excitation state $\ket{\Phi}$, while the inverse transition is strongly suppressed at
low temperature, $k_B T \ll \Omega$.
This mechanism does reflect on the evolution of populations
appearing in Eq.~\eqref{concxstate} leading to a faster increase
of the population term in $C_\rho^\Psi(t)$ than in
$C_\rho^\Phi(t)$, see Eqs.~(\ref{example1})~-~(\ref{example2}).
\begin{figure}
{\includegraphics[width=4.2 cm, height=2.6 cm]{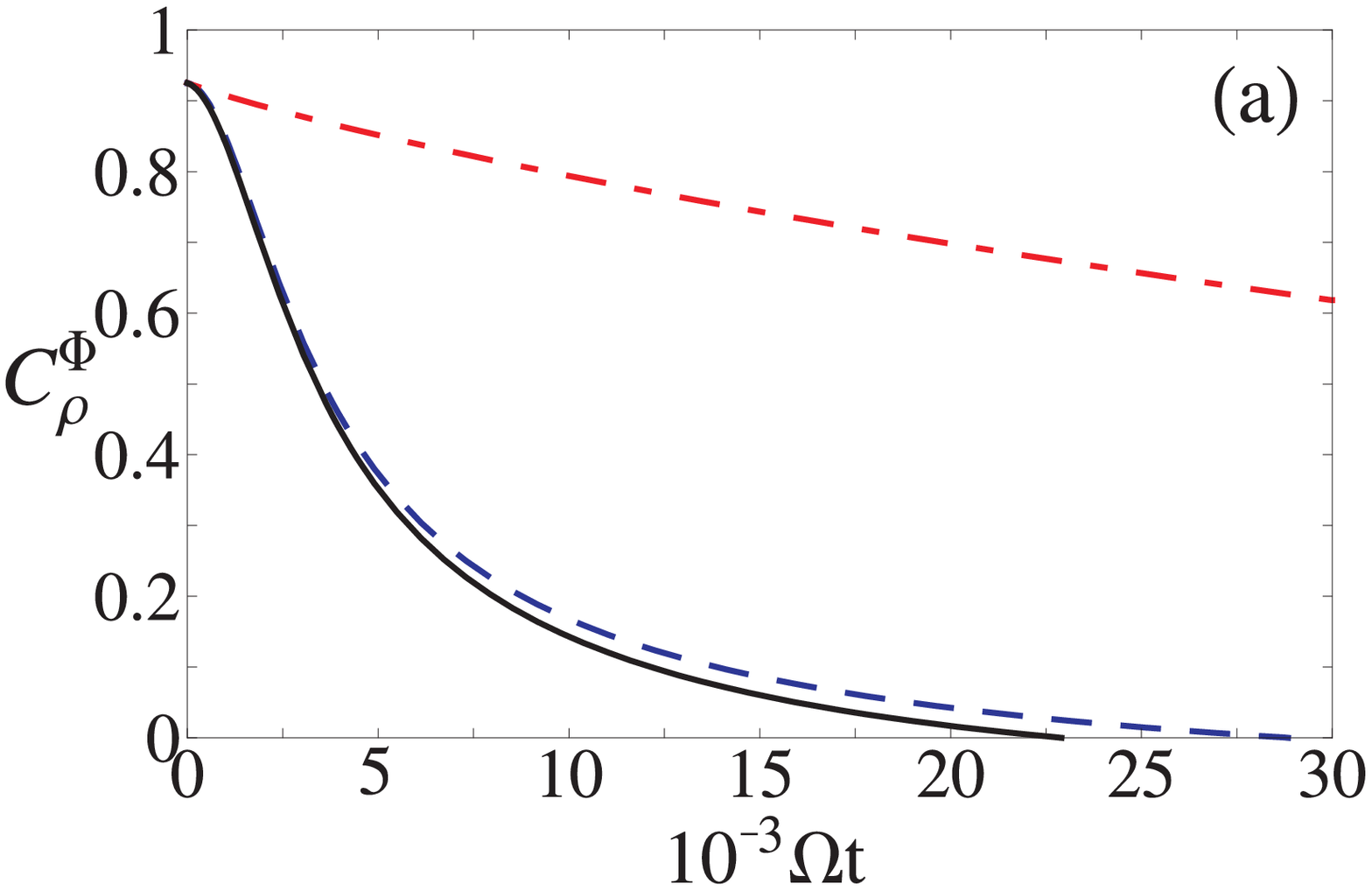}
\includegraphics[width=4.2 cm, height=2.6 cm]{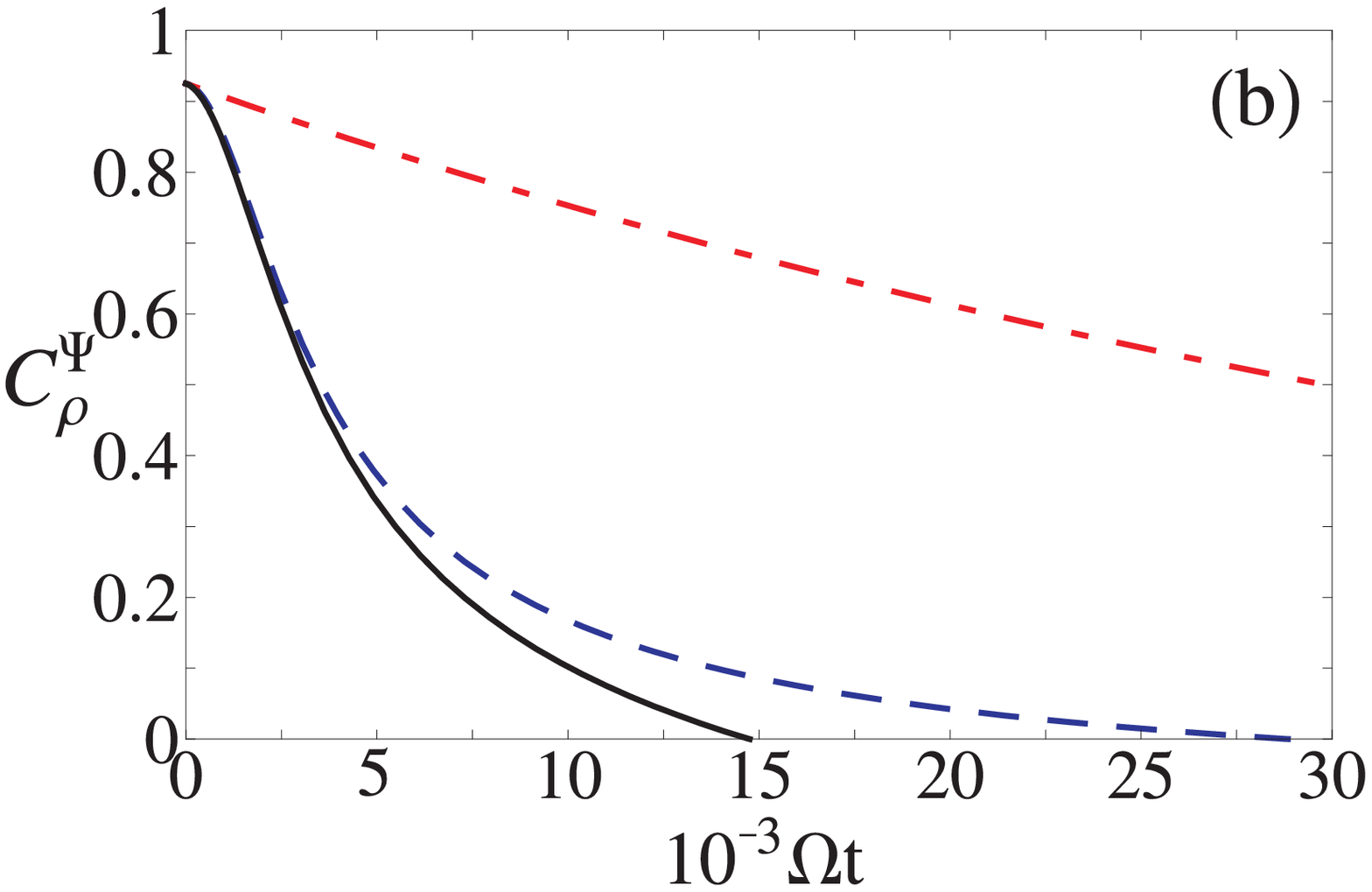}}
\caption{\label{fig:Cslowfast}\footnotesize (Color online)  $C_\rho^{\Phi}$
(panel a) and $C_\rho^{\Psi}$ (panel b) as a function of the
dimensionless time $\Omega t$ for $r=0.95$ with $a=1/\sqrt{2}$
for adiabatic (blue dashed), quantum (red dot-dashed), and their
interplay  (black solid). Noise and qubit characteristics as in Fig.~\ref{fig2}.}
\label{fig3}
\end{figure}
Note that for high purity levels, the ESD time due to quantum noise
is shorter than the adiabatic ESD time (Fig.~\ref{fig2} inset), which goes to infinity for
pure states. This observation however has to be supplemented
with a quantitative estimate of the amount of entanglement preserved before ESD takes place.
Indeed, for typical amplitudes of $1/f$ and white  noise, adiabatic noise considerably
reduces the amount of entanglement on a short time scale even for $r\to 1$. This is shown
in Fig.~\ref{fig:Cslowfast}, where $C_\rho^{\Phi}, (C_\rho^{\Psi}) \approx 1/\sqrt{2}$
(for lower values of concurrence Bell violation always occurs~\cite{verstraete2002PRL})
at $\Omega t \approx  2.38 \cdot 10^3, (2.24 \cdot 10^3)$.

{\em On-chip entanglement generation and maintenance  -- }
In the final part of this Communication we comment on the sensitivity of
the above analysis to experimental imperfections and on the
possibility to achieve, in a single chip, both entangling (preparation)
and local operations.
\begin{figure}
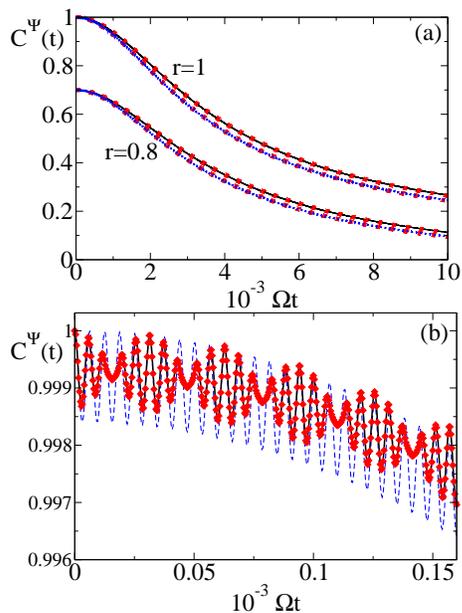

{\includegraphics[width=6 cm, height=4 cm]{BellomoFig4a.eps}
\includegraphics[width=6 cm, height=4 cm]{BellomoFig4b.eps}}
\caption{\label{fig4}\footnotesize (Color online)
$C_\rho^\Psi(t)$ as a function of the
dimensionless time $\Omega t$ for  $1/f$ noise
in $[1, 10^6]$~Hz, and $\Sigma/\Omega=0.02$ at $\theta=\pi/2$.
Panel (a): detuned qubits
 $\Omega_1 \, (\Omega_2) = 1, \, (1.2) \times  10^{11}$~rad/s (black),
or resonant qubits $\Omega=\Omega_1 $ (dashed blue).
Dotted lines are the SPA, Eq.~(\ref{conc-adiabatic}).
Panel (b): non-resonant qubits with (black) or without (dotted red (gray))
coupling  $\frac{g}{2} \sigma_{z,A} \otimes \sigma_{z,B}$, $g=10^9$~rad/s.
Uncoupled resonant qubits (dashed blue) from numerical simulations.
}
\end{figure}
A detuning between the two qubits of about  $20\%$
does not change even quantitatively our
analysis. This is illustrated in Fig.~\ref{fig4}(a) for different
initial purity of the Bell-like state $\rho^\Psi$. Note that the SPA is reliable even
when the dynamics of impurities responsible for $1/f$ noise extending up to
$\gamma_M \approx 10^6$~s$^{-1} \ll \Omega_\alpha$ is considered (numerical simulations).
Similar effects occur in the presence of a few per-cent deviations around
the fixed working point.

In addition, this picture is not modified if detuned qubits are coupled via
$-\frac{g}{2} \sigma_{z,A} \otimes \sigma_{z,B}$,  provided that
$g \ll |\Omega_1 - \Omega_2| , \Omega_\alpha$.
Deviations are visible only on a very small scale, Fig.~\ref{fig4}(b).
This fact points out that local (single qubit) operations may be performed with detuned qubits even
in the presence of a fixed transverse coupling (here we consider $\theta_\alpha=\pi/2$).
Tuning the individual qubit splittings on/off resonance effectively switches on/off their interaction.
This suggests that entanglement generation and local operations may be performed on-chip
in a fixed coupling scheme. Entanglement can be generated by tuning the qubits on resonance.
Without modifying the inter-qubit coupling, once induced a detuning,
entanglement may be maintained.
Both operations take place in the presence of broadband noise and their efficiency depends on 
the device operating point.
The present analysis shows that long~-~time maintenance of entanglement during local operations
can be achieved by operating the two qubits at their own optimal point.
In view of the different qualitative behavior of single-qubit and entanglement evolution,
this result was not as a priori established.
On the other side, it has been recently demonstrated that by properly fixing the qubits coupling
strength, high-fidelity two qubit operations can be
achieved and entanglement generated~\cite{preprint09}. It is therefore possible, depending on the
operation, to fix accordingly the optimal operating conditions of the nanodevice.
 In conclusion, we have studied a physical system where effectively amplitude and 
phase noise act on a X-state~\cite{yu2009Science}.
We demonstrated that even if adiabatic noise may  not 
induce ESD, for typical noise figures, it reduces the amount of entanglement initially
generated faster than quantum noise. The main effect of quantum noise consists in
differentiating classes of states more or less affected by relaxation processes. In the presence of
transverse noise, one-excitation states maintain
entanglement longer than two-excitation states.
Efficiency of entanglement preservation sensitively depends on the initial state purity.

\end{document}